\documentclass{INTERSPEECH2023}
\usepackage{comment}
\usepackage{amsmath}
\usepackage{makecell}
\DeclareMathOperator*{\argmax}{argmax}
\newcommand\Tstrut{\rule{0pt}{2.6ex}}   
\newcommand\Bstrut{\rule[-0.9ex]{0pt}{0pt}}
\usepackage[table,dvipsnames]{xcolor}

\ninept

\interspeechcameraready

\title{Sequence-Level Knowledge Distillation for Class-Incremental End-to-End Spoken Language Understanding}
\name{Umberto Cappellazzo$^1$, Muqiao Yang$^2$, Daniele Falavigna$^3$, Alessio Brutti$^3$}
\address{
  $^1$University of Trento, Trento, Italy\\
  $^2$Carnegie Mellon University, Pittsburgh, PA, USA \\
  $^3$Fondazione Bruno Kessler, Trento, Italy}
\email{umberto.cappellazzo@unitn.it, muqiaoy@andrew.cmu.edu, \{falavi,brutti\}@fbk.eu}

\begin{document}

\maketitle
 
\begin{abstract}
The ability to learn new concepts sequentially is a major weakness for modern neural networks, which hinders their use in non-stationary environments. Their propensity to fit the current data distribution to the detriment of the past acquired knowledge leads to the catastrophic forgetting issue.
In this work we tackle the problem of Spoken Language Understanding applied to a continual learning setting. We first define a class-incremental scenario for the SLURP dataset. Then, we propose three knowledge distillation (KD) approaches to mitigate forgetting for a sequence-to-sequence transformer model: the first KD method is applied to the encoder output (audio-KD), and the other two work on the decoder output, either directly on the token-level (tok-KD) or on the sequence-level (seq-KD) distributions. We show that the seq-KD substantially improves all the performance metrics, and its combination with the audio-KD further decreases the average WER and enhances the entity prediction metric.
\end{abstract}

\section{Introduction}
Spoken Language Understanding (SLU) is an essential component of any system that interacts with humans through speech, such as voice assistants and smart home devices \cite{tur2011spoken}. It is in charge of extrapolating the salient information from a spoken utterance so that proper actions can be taken to satisfy the user's requests. We can individuate two relevant tasks for SLU \cite{qin2021survey}: 1) Intent Classification, where we map the sentence to its corresponding intent, and 2) Entity Classification, or Slot Filling, by which we fill some fields of pre-defined semantic forms with content.
Traditional SLU systems \cite{mesnil2014using} employ a cascade of an automatic speech recognition (ASR) module followed by a natural language understanding module. Recently, end-to-end (E2E) strategies have garnered much attention \cite{lugosch2019speech,kim2021two} because they directly output semantic information from the audio, therefore reducing the impact of error propagation.

Most of the previous works on SLU have focused on the mainstream i.i.d. setting in which the entire dataset is available at once to the model \cite{seo2022integration,peng2023study}. However, this is in stark contrast with practical scenarios where models incur severe shifts in the data distribution or need to adapt to new domains without retraining from scratch. In such conditions, deep models tend to disrupt previous knowledge in favor of the fresh task, leading to catastrophic forgetting \cite{mccloskey1989}. This issue is tackled by the field of Continual Learning (CL) which endeavors to adapt a single model to learn a sequence of tasks such that the model performs properly both on new 
and prior 
tasks \cite{abraham2005memory}. Recently, multiple CL methods have been proposed, based on three main strategies \cite{de2021continual,parisi2019continual}: \textit{rehearsal-based} methods abate forgetting by retaining a portion of the old data \cite{lopez2017gradient}; \textit{regularization-based} approaches preserve the weights' importance through ad-hoc regularization loss terms \cite{li2017learning,ahn2021ss}, and \textit{architectural methods} modify the architecture of the model over time \cite{wang2022learning, yan2021dynamically}.

Recently, the challenging SLURP dataset~\cite{bastianelli2020slurp} has been released to 
address complex E2E SLU problems. In this paper, we propose to combine CL and SLU by defining a Class-Incremental Learning (CIL) setting for SLURP. Since each SLURP utterance is characterized by a domain scenario, we split the dataset into several tasks using the scenarios as a splitting criterion. Due to its lexical richness, the problems of intent and entity classification for SLURP are treated as a sequence-to-sequence (seq2seq) problem, where the intents and entities are generated along with the transcriptions. So, unlike the mainstream CL architecture composed of a feature extractor and a classifier, we exploit a transformer-based seq2seq architecture, whose decoder is also affected by forgetting as the encoder.

Our proposed approach combines rehearsal with regularization via knowledge distillation (KD) to combat forgetting at both the encoder and decoder levels. We investigate three KD approaches: one is applied to the encoder's output (audio-KD), whereas the other two distill the knowledge at the decoder side, either at a local (token-KD) or global (seq-KD) level. We show that the seq-KD stands out as the best approach, and we conduct a study where we integrate multiple KDs at once.


Our contributions can be summarized as follows: 1) We define a CIL scenario for the SLURP dataset, 2) we study how to moderate forgetting in a seq2seq model, thus moving away from the classical CL pipeline, and 3) we propose three KDs losses that effectively reduce forgetting and discuss their individual and combined contributions.

\section{Related work}
KD is a popular technique for model compression that allows transferring knowledge from a large, strong network, coined teacher, to a considerably smaller network, the student~\cite{hinton2015distilling,gou2021knowledge}. 
Besides the computer vision field, KD has also proved effective in NLP and speech-related tasks, where the student model mimics the teacher's distribution at a frame level. 
\cite{kim2016sequence} and \cite{takashima2018investigation}, for neural machine translation and CTC-based ASR, respectively, propose to apply the KD at a sequence level such that the student matches the probability distribution of the teacher's sequence obtained running beam search. 
More recently, \cite{choi2022temporal} advance an attention distillation method to transfer knowledge from a large transformer-based teacher by aligning its attention maps with those of the student through a Kullback-Leibler divergence loss.

The KD concept has also been exploited in CL. In this case, the teacher is the model trained in the previous tasks, whereas the student needs to be trained in the current task \cite{li2017learning}. The goal now is to transfer the knowledge of the old classes to the student model, which has no longer access to data related to them. In addition to the KD-based strategy, other kinds of CL strategies have been proposed in the speech domain. \cite{yang2022online} adapts the Gradient Episodic Memory method \cite{lopez2017gradient} to an online scenario for ASR. \cite{liu2022incremental} explores the use of the prompt-learning paradigm for class-incremental event detection. \cite{wang2022learning2} studies the use of self-supervised (SS) methods for continual representation learning for sound event classification, showing that SS learning is more resilient to forgetting than the supervised counterpart.

To the best of our knowledge, we are the first to explore how to attenuate forgetting in a seq2seq model for joint ASR/SLU. We define a CIL setting over the SLURP dataset, and we investigate the use of different KD methods applied to the encoder and decoder side. Our empirical evaluation shows the superiority of the sequence-level KD, and we elaborate on the entanglement between various KD combinations. 

\section{Class-Incremental learning for SLURP}
\label{sec:cil}
\begin{figure}[htb]
\centering
\includegraphics[width=4cm]{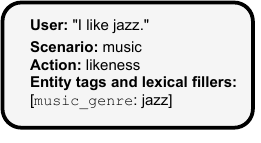}
\caption{Example of annotated utterance from the SLURP dataset. The intent in this case is the couple (music,likeness).}
\label{fig:utterance}
\end{figure}

In this section, we describe how we have defined the CIL setting for the SLURP dataset \cite{bastianelli2020slurp}. SLURP is a multi-domain dataset for E2E SLU comprising around 56 hours of audio of people interacting with a home assistant (\textit{slurp\_real}), with the addition of 43.5 hours of synthetic recordings (\textit{slurp\_synth}). At present this makes SLURP the biggest and the most diverse dataset in terms of lexical complexity for SLU. Each utterance is annotated with three semantics: Scenario, Action, and Entities. The pair (scenario, action) is defined as Intent. Overall, there are 18 unique scenarios, 46 actions (56 if we consider both \textit{slurp\_real} and \textit{slurp\_synth}), 55 entity types, and 69 intents. Figure~\ref{fig:utterance} provides an example of an annotated utterance.

\begin{figure}[htb]
\centering
\includegraphics[width=5.5cm]{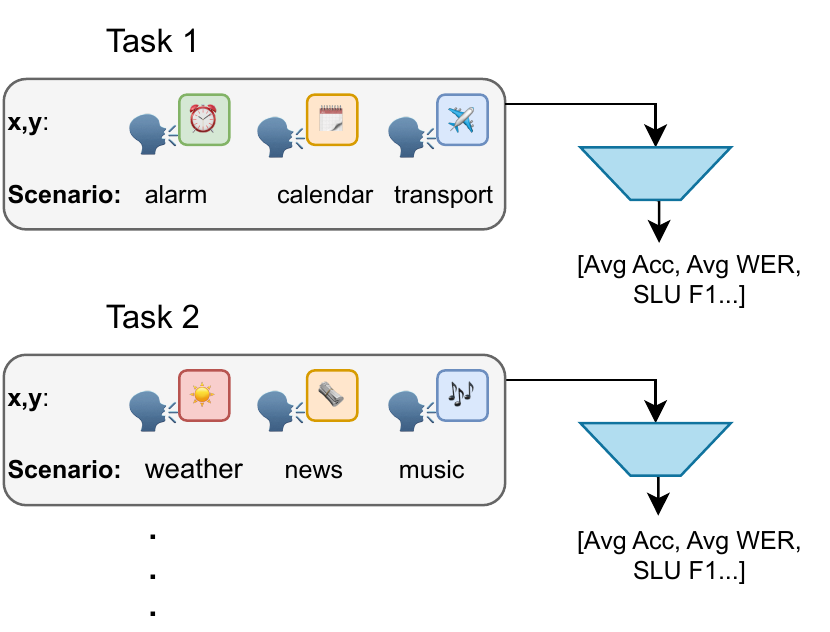}
\caption{The class-incremental learning setting for the SLURP dataset, where 3 new scenarios are introduced in each task.}

\label{fig:cl_scenario}
\end{figure}

We have used the scenarios as a splitting criterion to define the tasks of the CIL setting. The complete list of scenarios is: [``\textbf{alarm}'', ``\textbf{audio}", ``\textbf{calendar}'', ``\textbf{cooking}'', ``\textbf{datetime}'', ``\textbf{email}'', ``\textbf{general}'', ``\textbf{iot}'', ``\textbf{lists}'', ``\textbf{music}'', ``\textbf{news}'', ``\textbf{play}'', ``\textbf{qa}'', ``\textbf{recommendation}'', ``\textbf{social}'', ``\textbf{takeaway}'', ``\textbf{transport}'', ``\textbf{weather}'']. Since the number of scenarios is limited and each scenario provides a high-level concept associated with each utterance, we think that they can closely resemble a practical application that must adapt to new general domains. Additionally, since the intent classification is the chief metric to assess our model against, the use of scenarios as splitting criterion abides by the rule of having only intents related to scenarios available in the current task. Finally, although some actions and entities can be included in multiple scenarios, the overlap is very limited because the majority of the entities and actions are specific to a single scenario. For example, the action ``\textbf{taxi}'' is only associated with the scenario ``\textbf{transport}'', and the entity ``\textbf{weather\_descriptor}'' with the scenario ``\textbf{weather}''. Figure~\ref{fig:cl_scenario} shows two consecutive tasks, each introducing 3 new scenarios.

Another critical aspect is the order in which the scenarios are available to the model. In our implementation, the order depends on the cardinality so that the scenarios with the highest cardinality appear first.
In this way, we simulate a practical situation
in which we endow the model with the sufficient general knowledge, learning the largest scenarios first, that will be useful for learning more specific scenarios.

\section{Proposed approach}

As discussed in the previous section, we consider a CIL setting in which we want to adapt a single model to perform well on all seen tasks. Specifically, the training dataset is divided into $T$ distinct tasks, $\mathcal{D}=\{\mathcal{D}_0,\ldots,\mathcal{D}_{T-1}$\}, based on the scenario labels, so that a scenario is included in one and only one task. The dataset $\mathcal{D}_t$ of the $t^{th}$ task comprises audio signals $\mathcal{X}_t$ with associated transcriptions $\mathcal{Y}_t$, i.e. $\mathcal{D}_t=(\mathcal{X}_t,\mathcal{Y}_t)$. CIL setting is challenging as the model must be able to distinguish all classes till task $t$, thus at test time the task labels are not accessible (unlike task-incremental learning) \cite{hsu2018re}.

We employ a transformer-based seq2seq ASR architecture, constituted by a Wav2vec 2.0 encoder (WavEnc) \cite{baevski2020wav2vec} followed by a transformer decoder. Let $\textbf{x} = [x_1,\dots,x_I]$ be an audio input sequence of length $I$, and $\textbf{y} = [y_1,\dots,y_J]$ be the corresponding output sequence of length $J$, with $y_j \in \mathcal{V}$,
where $\mathcal{V}$ is the set of all possible output subword tokens. The goal of the ASR model is to find the most probable output sequence $\hat{\textbf{y}}$ given the input sequence $\textbf{x}$:
\begin{equation}
    \hat{\textbf{y}} = \argmax_{\textbf{y} \in \mathcal{Y}^*} p(\textbf{y}|\textbf{x};\theta),
\end{equation}
where $\mathcal{Y}^*$ is the set of all possible token sequences and $\theta$ represents the parameters of the seq2seq model.  

Suppose that $p(\textbf{y}|\textbf{x};\theta_{t})$ and $p(\textbf{y}|\textbf{x};\theta_{t-1})$ are the output probability distributions of the transformer decoder at task $t$ and $t-1$ parameterized by $\theta_{t}$ and $\theta_{t-1}$, respectively. The model at task $t-1$ can be seen as the teacher model. Let also $\mathcal{R}_{t}$ be the set of rehearsal data 
at the beginning of task $t$. In the following equations, we use $\textbf{x}\in\mathcal{D}_t$ in place of $(\textbf{x},\textbf{y})\in\mathcal{D}_t$ for brevity. The standard training criterion of rehearsal-based CL methods consists of minimizing the cross-entropy loss over $\mathcal{D}_t\cup\mathcal{R}_{t}$: 
\begin{equation}
    \mathcal{L}^t_{\text{CE}} =-\sum_{\textbf{x}\in\mathcal{D}_t\cup\mathcal{R}_{t}}\log(p(\textbf{y}|\textbf{x};\theta_{t})).
\label{eq:ce}
\end{equation}

The main idea of KD is to transfer knowledge from the teacher network $p(\textbf{y}|\textbf{x};\theta_{t-1})$ to a student model, such that the latter mimics the former's behavior. Basically, the KD is used to force the current model to not deviate too much from the teacher, which retains the knowledge of the previous tasks. We point out that the KD, unless otherwise stated, is applied to the sole rehearsal data since the teacher can effectively predict only the data seen in the previous tasks. 
We propose three different types of KDs: audio-KD, token-KD, and seq-KD. The audio-KD works at the encoder's output level, whereas the other two KDs are applied to the output of the decoder. In this way, we contrast forgetting either at the encoder or at the decoder side (or both, if we combine multiple KDs).

The {\bf audio-KD} forces the encoder's audio embeddings of the current task $t$ to resemble those from the previous task $t-1$. Let $\text{WavEnc}(\textbf{x}) \in \mathbb{R}^h$ be the Wav2vec 2.0 encoder output followed by a mean operation to squeeze the temporal dimension, where $h$ is the hidden size. We define the audio-KD loss as:

\begin{equation}
\mathcal{L}^t_{\text{audio-KD}} =\sum_{\textbf{x}\in\mathcal{R}_t}\Vert \text{WavEnc}_{\theta{_{t-1}}}(\textbf{x})-\text{WavEnc}_{\theta_{t}}(\textbf{x})\Vert^2,
\label{eq:kdaudio}
\end{equation}
where $\Vert \cdot \Vert$ is the Euclidean distance operator. Eq.~\ref{eq:kdaudio} acts as a regularization term for the encoder. 

We can apply such similar reasoning to the decoder, which predicts each word of the transcription in an autoregressive way (in our case we use Byte-Pair Encoding \cite{sennrich-etal-2016-neural}, so we will use the term token rather than word to refer to the output units). The {\bf token-KD} forces the current decoder to match the token-level distribution of the teacher. This is a kind of ``local" distillation in that the student mimics the teacher for each token of the transcription. The corresponding CE criterion is defined as:
\vspace{-0.15cm}
\begin{equation}
    \mathcal{L}^t_{\text{tok-KD}}=-\sum_{\textbf{x}\in \mathcal{R}_{t}}\sum_{j=1}^{J} p(y_j|\textbf{x},\textbf{y}_{<j};\theta_{t-1})\log(p(y_j|\textbf{x},\textbf{y}_{<j};\theta_{t}),
\label{eq:kdtoken}
\end{equation}
where $\textbf{y}_{<j}$ is the output sequence up to token j-1.

A potential flaw of this method is that if some initial token distributions are poorly estimated, their bias will be propagated until the end of the sequence. Indeed, a predicted token might be optimal at the current position in the sequence, but as we proceed through the rest of the sentence, it might turn out not to be the optimal one, given that later predicted positions are not already available.

{\bf Seq-KD} is an alternative approach that trains the student to generate the same output sequence as the teacher, thus working at the sequence level. 
In practice, we generate a new set of automatic transcriptions with the teacher model using beam search at the end of each task (``soft transcriptions"), and then we use them to train the student network with CE criterion in the next task. Formally, we add the following CE loss:
\begin{equation}
    \mathcal{L}^t_{\text{seq-KD}} =-\sum_{\textbf{x}\in \mathcal{R}_{t}}\log(p(\tilde{\textbf{y}}|\textbf{x};\theta_{t})),
\label{eq:seq}
\end{equation}
where $\tilde{\textbf{y}}$ is the output sequence generated with beam search using the teacher model. 

\begin{figure}[htb]
\centering
\includegraphics[scale=0.6]{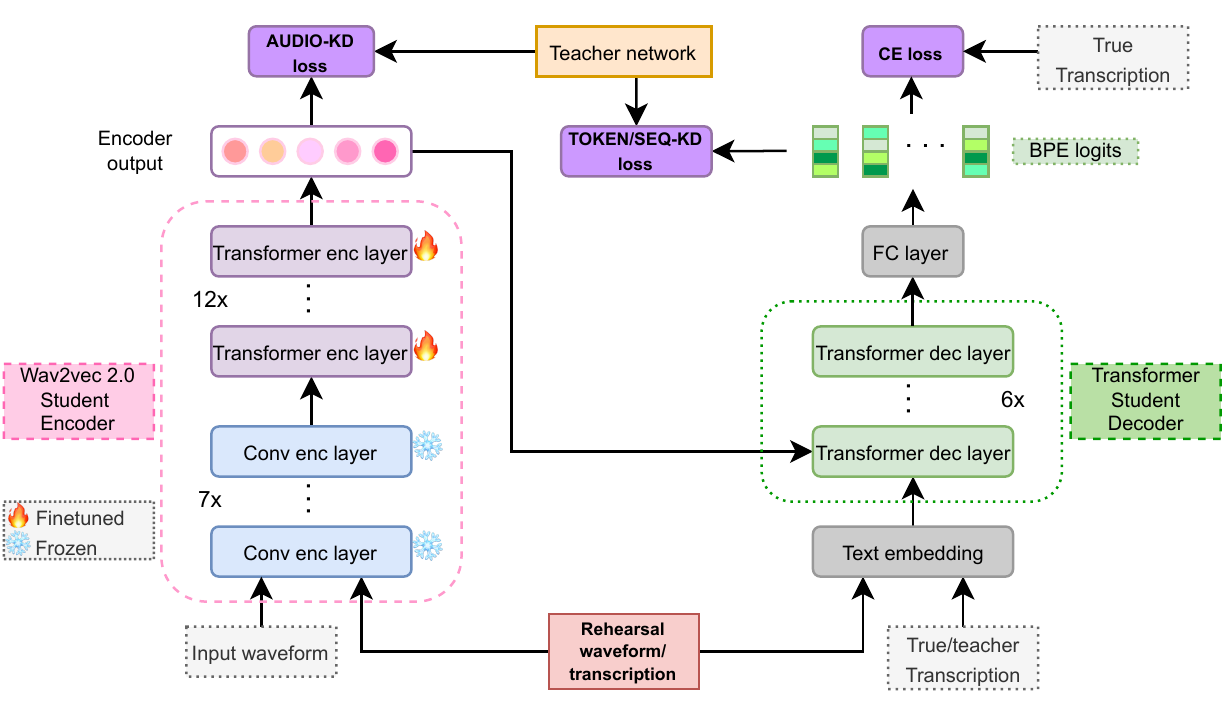}
\caption{Illustration of the learning process in the proposed CIL setting. The model from the current task (student) mimics the behavior of the teacher model through \textbf{audio}, \textbf{token}, and \textbf{sequence KD} losses to counter forgetting.}
\label{fig:model_pipeline}
\end{figure}

Overall, the total loss to be optimized 
at task $t$ is:

%
 \begin{equation}
 \mathcal{L}^t_{\text{TOT}}=(1 - \lambda_{\text{KD}})\mathcal{L}^t_{\text{CE}}+ \sum_{k \in \mathcal{K}} \lambda_{\text{KD}}\mathcal{L}^t_{k},
 \label{eq:tot}
\end{equation}
 where $\mathcal{K} = \{\text{audio-KD}, \text{tok-KD}, \text{seq-KD}\}$ and $\lambda_{\text{KD}}$ is a weighting parameter. Depending on whether we employ a single KD or multiple ones, Eq.~\ref{eq:tot} changes accordingly. Figure~\ref{fig:model_pipeline} shows the learning process with the three KD losses applied to the transformer architecture.
 


\section{Experiments}
\label{sec:experiments}

\begin{table*}[ht]
\centering
    \caption{Results 
    in terms of Average Accuracy ($\uparrow$), Last Accuracy ($\uparrow$), Average WER ($\downarrow$), and SLU F1 ($\uparrow$) for different strategies.}
    \begin{tabular}{lcccccccc}
    \toprule
    
      & & \multicolumn{2}{c}{\textbf{SLURP-3}\Tstrut} &  & & \multicolumn{2}{c}{\textbf{SLURP-6}\Tstrut} \\
     \textbf{Method}  & \multicolumn{4}{c}{-----------------------------------------} & \multicolumn{4}{c}{-----------------------------------------}\\

     & \textbf{Avg} & \textbf{Last} & \textbf{Avg}  &  & \textbf{Avg} & \textbf{Last} & \textbf{Avg} & \\ 
    &\textbf{Acc} &\textbf{Acc} & \textbf{WER} & \textbf{SLU F1} &\textbf{Acc} &\textbf{Acc} &\textbf{WER}  & \textbf{SLU F1}\\ 
    \hline 

    \textbf{Offline}\Tstrut\Bstrut & 85.84 & - & 20.46 & 70.59  & 85.84 & - & 20.46 & 70.59 \\
    \textbf{Fine-tuning} &46.27 &18.36 &35.82  &49.25 &33.56 &12.42 &46.26 &37.88 \\
    \hline
    \textbf{Rehe-5\% rand}\Tstrut &79.79 &74.82 &25.79  &65.85 &77.12 &73.11 &28.87 &63.22\\
    \textbf{Rehe-1\% rand} &71.30 &61.47 &29.13  &60.05 &66.11 &59.37 &34.77 &55.33\\
    \hline
    \textbf{Rehe-1\% iCaRL}\Tstrut &71.49 &61.66 &28.62  &60.23 &67.55 &62.55 &33.82 &56.09\\[0.1cm]
    \quad \textbf{+ audio-KD} &72.14 &63.03 &28.68  &61.08 &68.40 &62.83 &\textbf{32.04} &58.15\\[0.1cm]
    \quad \textbf{+ token-KD} &71.79 &61.54 &28.82  &\textbf{61.88} &68.36 &62.53 &32.47 &58.20\\[0.1cm]
    \rowcolor{CornflowerBlue!60}
    \quad \textbf{+ seq-KD} &\textbf{76.12} &\textbf{68.94} &\textbf{28.56}  &61.50 &\textbf{71.56} &\textbf{64.82} &32.50  &\textbf{58.29}\\
    \bottomrule

    \end{tabular}
    \label{tab:result}
\end{table*}
\subsection{Experimental settings}
\textbf{Dataset and CIL setting}.
We conduct experiments on the SLURP dataset \cite{bastianelli2020slurp} (see section~\ref{sec:cil}) using the official train, validation, and test splits, with a ratio of 70:10:20. In all experiments we also use \textit{slurp\_synth} only for training. Since very long audio data are harmful for efficient training, we remove the training samples longer than 7 seconds (around 0.004\% of the total training dataset).

Concerning the definition of the CIL setting, we experiment on two configurations: 1) the dataset is partitioned into 3 tasks, each comprising 6 scenarios (denoted as SLURP-3); 2) a more challenging configuration wherein the 18 scenarios are distributed across 6 tasks (denoted as SLURP-6). 

\textbf{Pre-processing and model configurations}.
 As proposed in \cite{arora2022espnet}, the intent and entity classification problems are treated as a sequence-to-sequence ASR task, where both intent and entities associated with an utterance are predicted alongside its transcription. In a sense, we build an ``augmented" transcription that will be fed to the transformer decoder, prepending the intent to the original transcription, followed by the entities and the corresponding lexical values. The special token \textit{\_SEP} is used to separate the intent from the entities and the entities from the original transcription, whereas the token \textit{\_FILL} is used to separate each entity from its value. If the original transcription is the one in Fig.~\ref{fig:utterance}, then the augmented transcription becomes: \textit{music\_likeness \_SEP music\_genre \_FILL jazz \_SEP I like jazz}.

{\bf Model.} The encoder is the base  Wav2vec 2.0 model pretrained and fine-tuned on 960 hours of Librispeech (a CNN-based feature extractor followed by 12 transformer blocks with hidden size = 768, 8 attention heads,  2048 FFN hidden states). The feature extractor is kept frozen during the training, whereas the transformer blocks are fine-tuned. Then, the transformer decoder includes 6 layers with the same parameters as the encoder. We apply layer normalization to the input raw waveforms. 
The total number of parameters of the model is around 148M. 

\textbf{Training}. We tokenize the transcriptions using Byte-Pair Encoding (BPE) \cite{sennrich-etal-2016-neural}, with a vocabulary size of 1k and BPE dropout = $0.1$. 
Both at inference time and for computing the soft labels for the KD-seq we run beam search with beam width = $20$. The number of epochs for each task is \{40,25,15\} for SLURP-3, whereas \{40,25,15,15,15,15\} for SLURP-6. The batch size is $32$. We use AdamW optimizer with learning rate = $5e^{-5}$ and weight decay = $0.1$. We use the validation set for hyperparameters tuning, and for selecting the best model for each task that is used for testing. 
Each experiment took approximately 1 day and a half on a Tesla V100 and a day on an Ampere A40. The code will be made public upon acceptance.

\textbf{CL baselines and strategies}.
Our upper bound is the {\it offline} method consisting of a single macro-task with all the scenarios (i.e. no incremental learning), while the naive {\it fine-tuning} approach, which retrains the same model task by task, is our lower bound. We consider two different sampling strategies for the rehearsal approach: 1) a random selection of the samples to retain, and 2) iCaRL \cite{rebuffi2017icarl}, which selects the samples closest to their moving barycenter. We provide an example with a memory buffer of size equal to around 5\% of the training dataset, and the rest of the experiments use 1\%. Finally, we show the result for each KD strategy, as well as their various combinations. The KD weight in Eq.~\ref{eq:tot} is proportional to the fraction of rehearsal data in the mini-batch and is defined as:
\begin{equation}
\lambda_{KD} =  
\sqrt{\frac{b_{rehe}}{b_{all}}},
\label{eq:weight}
\end{equation}

where $b_{rehe}$ is the number of rehearsal data in the current mini-batch, and $b_{all}$ is the current mini-batch size. We refer the reader to \cite{cappellazzo2022exploring} for a detailed description of this weight's choice.


\textbf{Metrics.} We evaluate the proposed methods using 4 metrics: the average intent accuracy, \textbf{Avg Acc}, 
after each task; the intent accuracy after the last task \textbf{Last Acc}; the average \textbf{SLU F1} metric for entity classification~\cite{bastianelli2020slurp}; the average word error rate, \textbf{Avg WER}, after each task.

\subsection{Results}

The performance for both CIL settings, SLURP-3 and SLURP-6, are reported in Table~\ref{tab:result}. First and foremost, we note that, as expected, the fine-tuning approach struggles in both settings, thus incurring catastrophic forgetting. The use of a rehearsal memory (rows Rehe-5\% and Rehe-1\%) proves to be very effective, even with only 1\% of retained data. Therefore, in the following experiments we consider 1\% of data in the rehearsal memory.
%
We also experiment with a more sophisticated sampling strategy, iCaRL \cite{rebuffi2017icarl}, which achieves noteworthy improvements, in particular for SLURP-6 (+1.44\% of Avg Acc).

When we focus on the proposed KDs, it is quite evident that the seq-KD leads to the most substantial improvement for both Avg and Last Acc metrics (+4.63\% and +7.28\% on SLURP-3). Instead, for WER and SLU F1, all three KDs behave similarly. Note that in our setting, previous intents are not seen anymore, and indeed the KDs help the model remember past scenarios. Conversely, though we expect the utterances to have some scenario-specific words, general speech tokens are spread among the tasks, making forgetting less critical for WER. Nevertheless, for the more challenging SLURP-6, KDs bring a notable enhancement also in terms of WER and SLU F1.  


\begin{figure}[htb]
\centering
\includegraphics[width=7cm]{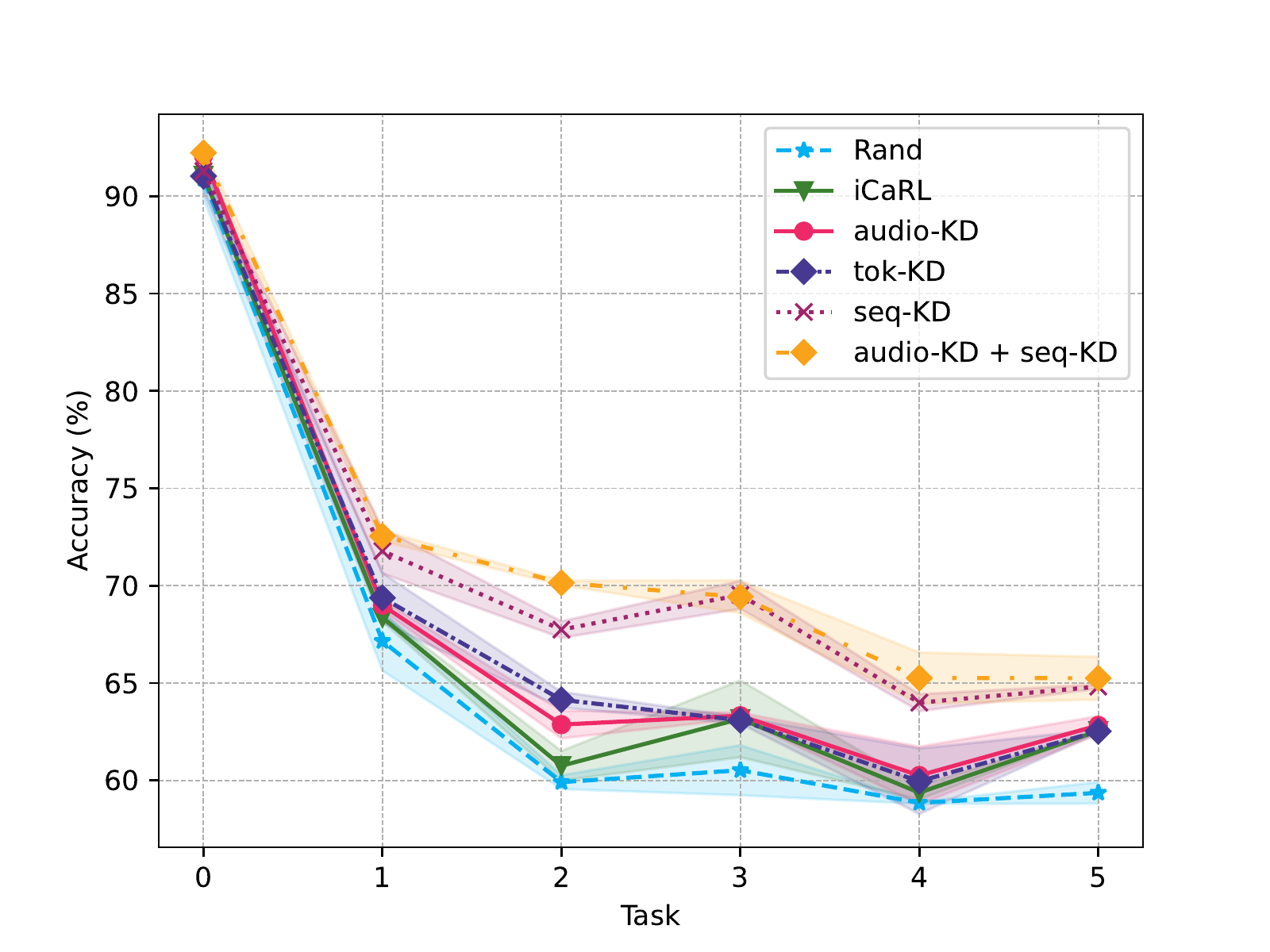}
\caption{The trend of the intent accuracy on the observed tasks for the SLURP-6 setting.}
\label{fig:avg_acc_trend}
\end{figure}

\textbf{Combination of multiple KDs.} In this final section, we investigate whether combining multiple KD approaches results in additional improvement. We try to combine two KDs at a time, and all three KDs together. The results for SLURP-6 are reported in Table~\ref{tab:result_comb}. As expected, the best combinations involve the use of seq-KD. Indeed, when the seq-KD is not included (audio + token), the results are even worse than using the KDs individually. Instead, the best combination is given by audio and seq KDs, the two approaches that yield the best improvement if taken singularly. We guess that
forcing the encoder output of the current task to be similar to that of the previous task (audio-KD) favors the cross-attention layer of the decoder to attend to the most relevant part of the audio signals. We also mention that using all three KDs leads to satisfactory results, yet slightly worse than seq + audio. We think that, for this last case, since four KDs are involved, the design of the KD weights is more cumbersome, and more experiments are necessary.

Finally, in Figure~\ref{fig:avg_acc_trend} we show the trend of the intent accuracy task by task for SLURP-6. We observe that the seq-KD outperforms both audio and token KDs by a large margin on all steps, and its combination with the audio KD leads to the best results.

As a last analysis, we point out that the additional computational burden brought by the proposed KDs is limited for two reasons: 1) the KD losses take into account only the rehearsal samples, which are just a fraction of the entire dataset (i.e., 1\%); 2) they just involve an additional forward pass through the teacher model, which is kept frozen during the current task. 

\begin{table}[ht]
\centering
    \caption{Analysis of different KD combinations on SLURP-6.}
    \begin{tabular}{lccccc}
    \Xhline{2\arrayrulewidth}

     \textbf{Combination}& \textbf{Avg} & \textbf{Last} & \textbf{Avg}  & \\ 
     &\textbf{Acc} &\textbf{Acc} & \textbf{WER} & \textbf{SLU F1} \\ 
    \hline 

    \textbf{audio + token}\Tstrut&68.13 &61.50 &32.46 &57.30 \\
     \rowcolor{CornflowerBlue!60}
     \textbf{audio + seq}&\textbf{72.48} &65.25 &\textbf{31.37}  &\textbf{60.00} \\
    \textbf{seq + token}&72.07 &63.46 &33.08 &58.25 \\
    \hline
    \textbf{audio + seq + token}\Tstrut &71.83 &\textbf{65.45} &32.55 &58.48 \\

    \Xhline{2\arrayrulewidth}
    \end{tabular}
    \label{tab:result_comb}
\end{table}

\vspace{-0.5cm}

\section{Conclusions}
In this paper, we define a CIL setting for a challenging SLU dataset, SLURP. To mitigate forgetting, we propose three different KD-based strategies working at different levels in the seq2seq model. Our extensive experiments reveal the superior performance of the seq-KD, and that combining multiple KDs results in additional improvements. In future work we will focus on refining the seq-KD by exploiting multiple beam search hypotheses with their corresponding scores, and we will carry out more experiments to find the optimal weights as multiple KDs are used. 

\bibliographystyle{IEEEtran}
\bibliography{mybib}

\begin{thebibliography}{10}
\providecommand{\url}[1]{#1}
\csname url@samestyle\endcsname
\providecommand{\newblock}{\relax}
\providecommand{\bibinfo}[2]{#2}
\providecommand{\BIBentrySTDinterwordspacing}{\spaceskip=0pt\relax}
\providecommand{\BIBentryALTinterwordstretchfactor}{4}
\providecommand{\BIBentryALTinterwordspacing}{\spaceskip=\fontdimen2\font plus
\BIBentryALTinterwordstretchfactor\fontdimen3\font minus
  \fontdimen4\font\relax}
\providecommand{\BIBforeignlanguage}[2]{{%
\expandafter\ifx\csname l@#1\endcsname\relax
\typeout{** WARNING: IEEEtran.bst: No hyphenation pattern has been}%
\typeout{** loaded for the language `#1'. Using the pattern for}%
\typeout{** the default language instead.}%
\else
\language=\csname l@#1\endcsname
\fi
#2}}
\providecommand{\BIBdecl}{\relax}
\BIBdecl

\bibitem{tur2011spoken}
G.~Tur and R.~De~Mori, \emph{Spoken language understanding: Systems for
  extracting semantic information from speech}.\hskip 1em plus 0.5em minus
  0.4em\relax John Wiley \& Sons, 2011.

\bibitem{qin2021survey}
L.~Qin, T.~Xie, W.~Che, and T.~Liu, ``A survey on spoken language
  understanding: Recent advances and new frontiers,'' \emph{IJCAI}, 2021.

\bibitem{mesnil2014using}
G.~Mesnil, Y.~Dauphin, K.~Yao, Y.~Bengio, L.~Deng \emph{et~al.}, ``Using
  recurrent neural networks for slot filling in spoken language
  understanding,'' \emph{IEEE/ACM TALSP}, vol.~23, no.~3, pp. 530--539, 2014.

\bibitem{lugosch2019speech}
L.~Lugosch, M.~Ravanelli, P.~Ignoto, V.~S. Tomar, and Y.~Bengio, ``Speech model
  pre-training for end-to-end spoken language understanding,'' \emph{Proc.
  Interspeech 2019}, pp. 814--818, 2019.

\bibitem{kim2021two}
S.~Kim, G.~Kim, S.~Shin, and S.~Lee, ``Two-stage textual knowledge distillation
  for end-to-end spoken language understanding,'' in \emph{ICASSP}, 2021, pp.
  7463--7467.

\bibitem{seo2022integration}
S.~Seo, D.~Kwak, and B.~Lee, ``Integration of pre-trained networks with
  continuous token interface for end-to-end spoken language understanding,'' in
  \emph{ICASSP}, 2022, pp. 7152--7156.

\bibitem{peng2023study}
Y.~Peng, S.~Arora, Y.~Higuchi, Y.~Ueda, S.~Kumar, K.~Ganesan, S.~Dalmia,
  X.~Chang, and S.~Watanabe, ``A study on the integration of pre-trained ssl,
  asr, lm and slu models for spoken language understanding,'' in \emph{SLT},
  2023, pp. 406--413.

\bibitem{mccloskey1989}
M.~McCloskey and N.~J. Cohen, ``Catastrophic interference in connectionist
  networks: The sequential learning problem,'' in \emph{Psychology of learning
  and motivation}.\hskip 1em plus 0.5em minus 0.4em\relax Elsevier, 1989,
  vol.~24, pp. 109--165.

\bibitem{abraham2005memory}
W.~C. Abraham and A.~Robins, ``Memory retention--the synaptic stability versus
  plasticity dilemma,'' \emph{Trends in neurosciences}, vol.~28, no.~2, pp.
  73--78, 2005.

\bibitem{de2021continual}
M.~De~Lange, R.~Aljundi, M.~Masana \emph{et~al.}, ``A continual learning
  survey: Defying forgetting in classification tasks,'' \emph{TPAMI}, vol.~44,
  no.~7, pp. 3366--3385, 2021.

\bibitem{parisi2019continual}
G.~I. Parisi, R.~Kemker, J.~L. Part, C.~Kanan, and S.~Wermter, ``Continual
  lifelong learning with neural networks: A review,'' \emph{Neural Networks},
  vol. 113, pp. 54--71, 2019.

\bibitem{lopez2017gradient}
D.~Lopez-Paz and M.~Ranzato, ``Gradient episodic memory for continual
  learning,'' \emph{NeurIPS}, vol.~30, 2017.

\bibitem{li2017learning}
Z.~Li and D.~Hoiem, ``Learning without forgetting,'' \emph{IEEE transactions on
  pattern analysis and machine intelligence}, vol.~40, no.~12, pp. 2935--2947,
  2017.

\bibitem{ahn2021ss}
H.~Ahn, J.~Kwak, S.~Lim, H.~Bang, H.~Kim, and T.~Moon, ``Ss-il: Separated
  softmax for incremental learning,'' in \emph{ICCV}, 2021, pp. 844--853.

\bibitem{wang2022learning}
Z.~Wang, Z.~Zhang, C.-Y. Lee, H.~Zhang, R.~Sun, X.~Ren, G.~Su, V.~Perot, J.~Dy,
  and T.~Pfister, ``Learning to prompt for continual learning,'' in
  \emph{Proceedings of CVPR}, 2022, pp. 139--149.

\bibitem{yan2021dynamically}
S.~Yan, J.~Xie, and X.~He, ``Der: Dynamically expandable representation for
  class incremental learning,'' in \emph{CVPR}, 2021, pp. 3014--3023.

\bibitem{bastianelli2020slurp}
E.~Bastianelli, A.~Vanzo, P.~Swietojanski, and V.~Rieser, ``Slurp: A spoken
  language understanding resource package,'' \emph{arXiv preprint
  arXiv:2011.13205}, 2020.

\bibitem{hinton2015distilling}
G.~Hinton, O.~Vinyals, and J.~Dean, ``Distilling the knowledge in a neural
  network,'' \emph{arXiv preprint arXiv:1503.02531}, 2015.

\bibitem{gou2021knowledge}
J.~Gou, B.~Yu, S.~J. Maybank, and D.~Tao, ``Knowledge distillation: A survey,''
  \emph{International Journal of Computer Vision}, vol. 129, pp. 1789--1819,
  2021.

\bibitem{kim2016sequence}
Y.~Kim and A.~M. Rush, ``Sequence-level knowledge distillation,'' in
  \emph{Proceedings of the 2016 Conference on Empirical Methods in Natural
  Language Processing}, 2016, pp. 1317--1327.

\bibitem{takashima2018investigation}
R.~Takashima, S.~Li, and H.~Kawai, ``An investigation of a knowledge
  distillation method for ctc acoustic models,'' in \emph{2018 IEEE
  International Conference on Acoustics, Speech and Signal Processing
  (ICASSP)}, 2018, pp. 5809--5813.

\bibitem{choi2022temporal}
K.~Choi, M.~Kersner, J.~Morton, and B.~Chang, ``Temporal knowledge distillation
  for on-device audio classification,'' in \emph{ICASSP}, 2022, pp. 486--490.

\bibitem{yang2022online}
M.~Yang, I.~Lane, and S.~Watanabe, ``Online continual learning of end-to-end
  speech recognition models,'' \emph{InterSpeech}, 2022.

\bibitem{liu2022incremental}
M.~Liu, S.~Chang, and L.~Huang, ``Incremental prompting: Episodic memory prompt
  for lifelong event detection,'' in \emph{Proceedings of the 29th
  International Conference on Computational Linguistics}, 2022, pp. 2157--2165.

\bibitem{wang2022learning2}
Z.~Wang, C.~Subakan, X.~Jiang, J.~Wu, E.~Tzinis, M.~Ravanelli, and
  P.~Smaragdis, ``Learning representations for new sound classes with continual
  self-supervised learning,'' \emph{IEEE Signal Processing Letters}, vol.~29,
  pp. 2607--2611, 2022.

\bibitem{hsu2018re}
Y.-C. Hsu, Y.-C. Liu, A.~Ramasamy, and Z.~Kira, ``Re-evaluating continual
  learning scenarios: A categorization and case for strong baselines,''
  \emph{arXiv preprint arXiv:1810.12488}, 2018.

\bibitem{baevski2020wav2vec}
A.~Baevski, Y.~Zhou, A.~Mohamed, and M.~Auli, ``wav2vec 2.0: A framework for
  self-supervised learning of speech representations,'' \emph{NeurIPS},
  vol.~33, pp. 12\,449--12\,460, 2020.

\bibitem{sennrich-etal-2016-neural}
R.~Sennrich, B.~Haddow, and A.~Birch, ``Neural machine translation of rare
  words with subword units,'' in \emph{Proceedings of the 54th Annual Meeting
  of the Association for Computational Linguistics}, 2016, pp. 1715--1725.

\bibitem{arora2022espnet}
S.~Arora, S.~Dalmia, P.~Denisov, X.~Chang \emph{et~al.}, ``Espnet-slu:
  Advancing spoken language understanding through espnet,'' in \emph{ICASSP},
  2022, pp. 7167--7171.

\bibitem{rebuffi2017icarl}
S.-A. Rebuffi, A.~Kolesnikov, G.~Sperl, and C.~H. Lampert, ``icarl: Incremental
  classifier and representation learning,'' in \emph{CVPR}, 2017, pp.
  2001--2010.

\bibitem{cappellazzo2022exploring}
U.~Cappellazzo, D.~Falavigna, and A.~Brutti, ``Exploring the joint use of
  rehearsal and knowledge distillation in continual learning for spoken
  language understanding,'' \emph{arXiv preprint arXiv:2211.08161}, 2022.

\end{thebibliography}

\end{document}